\documentclass[10pt, twocolumn]{IEEEtran}
\usepackage{filecontents}
\usepackage{amsthm}
\usepackage[fleqn]{amsmath}
\usepackage[table,xcdraw]{xcolor}

\usepackage{epsfig}
\usepackage{times}
\usepackage{verbatim}
\usepackage{enumerate}
\usepackage{multicol,afterpage,wrapfig}
\usepackage{booktabs}
\usepackage{graphicx}
\usepackage{ifmtarg}
\usepackage{amssymb}
\usepackage{amsmath}
\usepackage{xcolor}
\usepackage{bbm}
\usepackage{cite}
\usepackage{epstopdf}
\usepackage{subcaption}
\usepackage{nicefrac}

\usepackage{caption}
 
\DeclareMathOperator*{\argmax}{arg\,max}
\DeclareMathOperator*{\argmin}{arg\,min}

\def\BibTeX{{\rm B\kern-.05em{\sc i\kern-.025em b}\kern-.08em
    T\kern-.1667em\lower.7ex\hbox{E}\kern-.125emX}}

\begin{document}
\title{Grip-Aware Analog mmWave Beam Codebook Adaptation for 5G Mobile Handsets}
 \author{
 Ahmad AlAmmouri, Jianhua Mo, Boon Loong Ng, Jianzhong Charlie Zhang,  and Jeffrey G. Andrews\\ \thanks{A. AlAmmouri and J. G. Andrews are with The Wireless Networking and Communications Group (WNCG), The University of Texas at Austin, Austin, TX 78712 USA.}
 \thanks{J. Mo, B. L. Ng, and J. C. Zhang are with Samsung Research America: Standards and Mobility Innovation Lab, Plano, TX 75023 USA.} \thanks{This is part of our work in \cite{Hand_AlAmmouri19}. Hence, for more results, explanations, and discussions, refer to \cite{Hand_AlAmmouri19}.}
 \vspace{-0.8cm}}

\maketitle
\thispagestyle{plain}
\pagestyle{plain}

	\begin{abstract}
	This paper studies the effect of the user hand grip on the design of beamforming codebooks for 5G millimeter-wave (mmWave) mobile handsets. The high-frequency structure simulator (HFSS) is used to characterize the radiation fields for  fourteen possible handgrip profiles based on experiments we conducted. The loss from hand blockage on the antenna gains can be up to $20-25$ dB, which implies that the possible hand grip profiles need to be taken into account while designing  beam codebooks.  Specifically, we consider three different codebook adaption schemes: a grip-aware scheme, where perfect knowledge of the hand grip is available; a semi-aware scheme, where just  the application (voice call, messaging, etc.) and the orientation of the mobile handset is known; and a grip-agnostic scheme, where the codebook ignores hand blockage.  Our results show that the ideal grip-aware scheme can provide more than $50\%$ gain in terms of the spherical coverage over the agnostic scheme, depending on the grip and orientation. Encouragingly, the more practical semi-aware scheme we propose provides performance approaching the fully grip-aware scheme.  Overall, we demonstrate that 5G mmWave handsets are different from pre-5G handsets: the user grip needs to be explicitly factored into the  codebook design.

	\end{abstract}

	\section{Introduction}

   A critical aspect of millimeter-wave (mmWave) signals  is its sensitivity to blockage by the human hand \cite{Statistical_Raghavan18} which has to be accounted for while designing beamforming codebooks. Beamforming, which is required to combine the signals from different antennas, is a critical part of mmWave communications since it relies on high array gains \cite{Millimeter_Raghavan17}. 
   
   Although digital beamforming is elegant theoretically and results in  excellent array gains, the commercial devices will most probably avoid using fully-digital beamforming due to its complexity, both in hardware and signal processing \cite{MIMO_Alkhateeb14, Automatic_Jianhua19}. Instead, analog beamforming, which is based on pre-designed codebooks, is proposed as an alternative and supported by the 3GPP 5G NR standard \cite{3GPP2018}. 
   The idea is that a set of beamforming vectors (codewords) is predetermined, and the device switches between these codewords to select the one that maximizes the antenna gain in the direction of the received signal. A comparison between digital and analog codebook beamforming has been conducted recently in \cite{Antenna_Raghavan19}, where the authors showed that the codebook beamforming can balance the trade-off between the overhead complexity and array gain and can perform close to the digital beamforming. Another recent study was performed in \cite{Automatic_Jianhua19}, which proposes and compares different heuristic approaches to design the beam codebooks taking into account the antennas' radiation pattern. 
	
	However, \cite{Automatic_Jianhua19} only considers the free space propagation without the possible hand blockage and \cite{Antenna_Raghavan19} focuses on comparing different antenna types and comparing the analog codebook beamforming and the digital one. Moreover, their model of the human hand is either based on a stochastic model \cite{3GPP2017}, which assumes a 30 dB flat loss in the antenna gain due to the hand blockage, or a statistical loss \cite{Statistical_Raghavan18}, which only differentiates between the blockage in landscape and portrait orientations. It is well-known that users hold their phones in different ways depending on the applications they are using, the environment,  and their habits. These different hand grips affect the radiation patterns in distinct ways even though the phone orientation is the same. For example, while in the landscape orientation, the user can hold the phone with both hands or a single hand. The effect of single hand blockage on the radiation patterns can be quite different from that of dual hand blockage because when the fingers are near the antenna, the radiation is influenced more by coupling, reflection, and attenuation caused by the finger \cite{mmWave_Zhao15,Efficiency_Khan18}.  
	
	In this work, we focus on the problem of adapting the beam codebooks according to the available knowledge of the user hand grip. More specifically, we consider a practical placement of the antennas on the mobile device and a practical design algorithm of the beam codebooks. We also use high-frequency structure simulator (HFSS) to obtain the antenna radiation patterns, where we include our model of the human hand into the simulations to capture the irregularities in the radiation patterns caused by the hand and to avoid using oversimplified analytical models. To this end, we compare three codebook adaptation schemes: An idealistic {\it grip-aware} scheme, which assumes that the device can accurately detect the hand grip and use a codebook that is specifically designed for it; a more practical scheme we call the {\it semi-aware} scheme, which is based on the assumption that the device only knows the orientation of the device and the active application the user is using; and a benchmark {\it grip-agnostic} scheme, where the codebook is designed assuming no-blockage and does not accommodate the hand grip. We show that the {\it grip-aware} and the {\it semi-aware} schemes can achieve over $57\%$ gain in terms of the spherical coverage compared to the {\it grip-agnostic} scheme, where the gain depends on the activity. Overall, our results show that beamforming codebooks must be adapted to the hand grip. Otherwise, a significant loss of the spherical coverage is expected which might lead to a link failure.
	
	\section{System Model}\label{Sec:SysMod}

	\subsection{Antenna and Mobile Design}

\begin{figure}[t]
		\centering
			\centerline{\includegraphics[width=  3in]{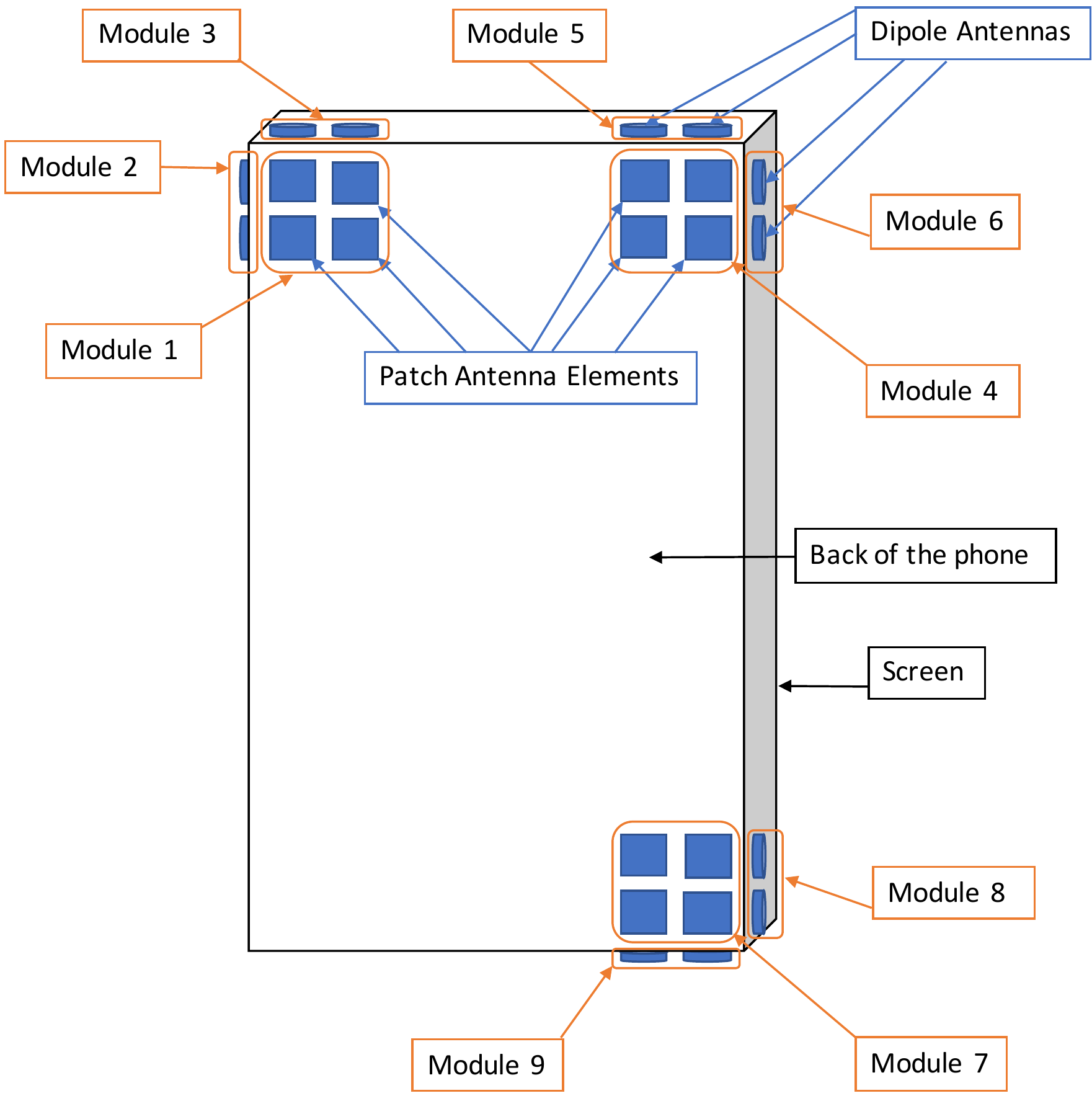}}
		\caption{\, The considered antenna placements.}
		\label{fig:Designs}
	\end{figure}

	 A good design for the mobile device must have different antenna arrays with different orientations, such that each array is able to receive the signal from certain directions (coverage region). What's more, there must be overlap between the different coverage regions for the antenna arrays to account for possible hand blockage. To this end, different antenna designs were proposed in 3GPP meetings  \cite{Samsung_Consideration17,Qualcomm_Consideration17,Apple_Consideration17} and studied in the literature \cite{Antenna_Raghavan19} to ensure the previous requirements. We consider the design shown in Fig. \ref{fig:Designs}, where we have three $2\times 2$ patch antenna modules placed on top two corners and the right bottom corner of the back of the phone. Also, each is surrounded by two modules of dipole antennas on the edges. 
	Note that we consider a mix of edge (dipole) and face (patch) antenna arrays. This is because the $2\times2$ patch array provides good spherical coverage to the hemisphere on the back of the phone, but it cannot be used to receive (transmit) signals from (through) the front end of the phone because of the front-to-back ratio of the patch antenna as well as blockage by the screen \cite{Statistical_Raghavan18}. On the other hand, the dipole antenna arrays can be used to extend the coverage region to the sides of the device, but it also restricts us to linear antenna arrays due to size constraints. Note that there is also redundancy since some arrays point to the same direction. This is because we need to account for possible hand blockages which can severely reduce the gain of the blocked antenna as shown later in this work.
    
    To reduce the complexity of the management and design of the beam codebook, the mobile device is restricted to use one array at a time, referred to as an antenna module, which is either a $2 \times 2$ patch array or a $1 \times 2$ dipole array. This design will be studied in the following sections in terms of the spherical coverage in free space as well as in the presence of hand blockage.
	\subsection{Spherical Coverage and HFSS}\label{Sec:SCaH}
	
	The main performance metric considered in this work is {\it spherical coverage} which corresponds to the antennas' far field gain over all possible directions. To quantify it, we discretize the unit sphere uniformly, and then the antenna gain is found for each point on the sphere. More rigorously, let $\mathcal X=\left\{x_i=(\theta_i,\phi_i), \ \forall i \in \{1,2, \cdots, N_p\}\right\}$ be a set of points uniformly distributed on the surface of the unit sphere, where $N_p$ is the number of points 
	and  $0^\circ \leq \theta_i \leq 180^\circ $, $0^\circ \leq \phi_i < 360^\circ$ are the spherical coordinates of the $i^{\rm th}$ point. Each point represents one possible direction for the signal to arrive from. In other words, we have discretized the possible angle of arrivals (AoAs). The handset has a set of beamforming vectors (codewords), referred to as a codebook $\mathcal W_c$, and chooses the codeword that maximizes the antenna gain given the AoA. Let the gain at point $x_i$ be denoted by $G_{i}({\mathbf w})\in \mathbb{R}_{+}$, given a codeword ${\mathbf{w}}\in \mathbb{C}^{N_t \times 1}$, where $N_t$ is the total number of antennas.  Then the spherical coverage is defined as follows
	 \begin{equation}\label{eq:SphericalCovMat}
	\mathcal{S}(\mathcal{W}_c)=\{\max_{{\mathbf w} \in {\mathcal W_c}} G_i({\mathbf w}), \ \forall i \in \{1,2, \cdots, N_p\}\}.
	\end{equation}
	
	Based on this, the mean spherical coverage across all points on the sphere given a codebook $\mathcal W_c$, denoted by $\bar{\mathcal{S}}(\mathcal W_c)$,  is given by
	\begin{equation}\label{Eq:AveSC}
	\bar{\mathcal{S}}({\mathcal W_c})=\frac{1}{N_p} \sum\limits_{i=1}^{N_p} \max_{\mathbf w \in {\mathcal W_c}} G_i(\mathbf w).
	\end{equation}

	To compute the antenna gain $G_{i}({\mathbf w})$ given a certain setup of antennas, we use an electromagnetic simulation software called HFSS, since the theoretical models fall short in capturing the irregularities in the antennas' radiation patterns. The design shown in Fig. \ref{fig:Designs} was built in HFSS, where the antennas were designed for 39 GHz carrier frequency with a half wavelength separation. In this design, we did not add the different components in the phone and only modeled the screen, since it is the main source of reflections and absorption.

	Using HFSS, we get the antenna response of each element in each array for each direction $x_i$. Hence, for a given direction $x_i$, we have an antenna response vector, which we denote by ${\mathbf{M}_{i}}\in \mathbb{C}^{1 \times N_t}$, and the antenna gain is given by
	\begin{equation}\label{Eq:Gain}
	G_{i}({\mathbf w})= {\mathbf{w}}^{H} {\mathbf{M}}_{ i}^{H} {\mathbf{M}}_{i} {\mathbf{w}},
	\end{equation}
	where ${\mathbf w}^{H}$ is the Hermitian (conjugate) transpose of the vector ${\mathbf w}$.

 \begin{figure*}
	\centering
	\begin{subfigure}{\columnwidth}
		\centerline{\includegraphics[width=  2.6in]{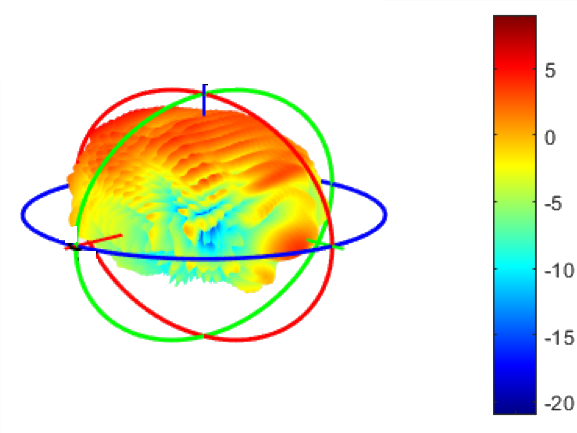}}
		\caption{\, Without blockage.}
		\label{fig:RadWO}
	\end{subfigure}%
	\begin{subfigure}{\columnwidth}
		\centering
		\centerline{\includegraphics[width=  2.6in]{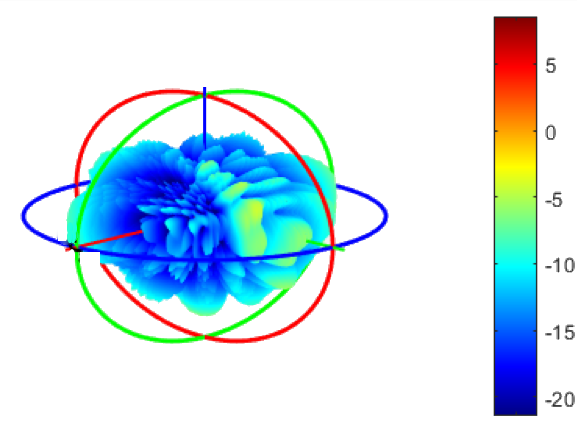}}
		\caption{\, With blockage.}
		\label{fig:RadW}
	\end{subfigure}
	\caption{The 3D antenna gain in dB of a patch antenna with and without blockage by a finger. }
	\label{fig:RadWOandW}\vspace{-0.4cm}
\end{figure*}

	At this point, we need to design the codebook  to maximize some function of the spherical coverage.
	In this work, we choose the codebook ${\mathcal W_c}$ that maximizes the mean spherical coverage out of a given large set of candidate codewords ${\mathcal W_d}$, where the set ${\mathcal W_d}$ satisfies all the required conditions on the maximum transmit power and the finite number of bits in the phase shifter. Hence, we have the following optimization problem
	\begin{equation}
	{\mathcal W_{c}}=\argmax_{ \left\{ {\mathbf w_1}, \cdots, {\mathbf w}_{N_c} \right\} \subset {\mathcal W_d}} \bar{\mathcal{S}}(\{{{\mathbf w}_1, \cdots, {\mathbf w}_{N_c}}\}),\label{Eq:OptiPro}
	\end{equation}
	which is a combinatorial problem and solving it using an exhaustive search is not feasible given a large set ${\mathcal W_d}$ and a large number of points $N_p$. Different heuristic approaches to solve this problem were discussed in detail in \cite{Automatic_Jianhua19}. In this work, we adopt the greedy approach which we now  describe. The greedy approach is an iterative one, where the first codeword is selected to maximize the mean spherical coverage given by \eqref{Eq:AveSC}. The second codeword is also chosen to maximize the mean spherical coverage, but given that we have already selected the first codeword. In other words, the second codeword is chosen such that the composite spherical coverage of the first and second codewords is maximized. Hence, the choice of second codeword is biased to select the one that points the beam in a different direction of the first codeword. Subsequently,  the $N^{\rm th}$ codeword is found by
	  \begin{align}
	{\mathbf w}_{N}=\argmax_{{\mathbf w}\in {\mathcal W_d} \setminus \{{\mathbf w}_1, \cdots,{\mathbf w}_{N-1}\}} \frac{1}{N_p}f({\mathbf w}),
	\end{align}
	where
	\begin{align}
	&\!\!\!\!\!\!\!\!\!f({\mathbf w})=  \sum\limits_{i=1}^{N_p}  \max \left( G_i({\mathbf w}),G_i({\mathbf w}_1), \cdots, G_i({\mathbf w}_{N-1}) \right).\label{Eq:OptiProN}
	 \end{align}

	This approach is only optimal if the desired codebook size is one and otherwise, it is suboptimal. However, it was shown in \cite{Automatic_Jianhua19} that it performs  well compared to other more sophisticated algorithms.
	
	Before wrapping up this section, we discuss the choice of the candidate set of codewords ${\mathcal W_d}$. There are many options to construct ${\mathcal W_d}$ as shown in \cite{Automatic_Jianhua19}. In this work, we select a subset of points $\mathcal Y$ from $\mathcal X$ (i.e., $\mathcal Y \subset \mathcal X$) that has a length $N_d \ll N_p$. Then, for each point $y\in \mathcal Y$, we select the codeword that maximizes the antenna gain at this point by eigenvalue decomposition. Finally, the phases of the codewords are quantized according to the number of bits of the phase shifters $N_b$.  Note that the choice of ${\mathcal W_d}$ is not very critical as shown in \cite{Automatic_Jianhua19} as long as $N_d$ is big enough. 
	
	\subsection{Hand Model}
	
	There are several works in the literature that study the effect of the hand blockage on the antennas' radiation patterns in mmWave bands. For quantitative and analytical models, the 3GPP standard \cite{3GPP2017} assumes a 30 dB loss across a region on the sphere around the cell phone, where the boundaries of this region depend on whether the phone is held in landscape or portrait orientation. This model was revisited in \cite{Statistical_Raghavan18}, where a statistical model for the affected region was proposed instead of a flat 30 dB loss. However, both of these models do not differentiate between different hand grips and only distinguish the portrait and the landscape orientations of the phone.

	    As we mentioned earlier, different hand grips affect the radiation patterns in distinct ways although the phone orientation is the same. Consequently, we cannot rely on the models used in \cite{3GPP2017,Statistical_Raghavan18}.
		Hence, our approach is to model the human fingers and then include it into HFSS simulations which will take care of all the possible irregular effects of the hand on the antenna radiation pattern. More specifically, 
	we consider a two-layer model, where the first corresponds to the skin and the second models the rest of the human finger. The dielectric properties for the skin are averaged over the wet and dry conditions and the dielectric properties for the second layer are averaged over the properties of the bone, muscle, and fat. The used values are shown in Table 1 and taken from     \cite{An_Andreuccetti12}. The size of the finger is taken to be the average size of personnel in the US Army as recorded in \cite{Hand_Greiner91} and the thickness of the skin is assumed to be $2$ mm \cite{mmWave_Zhao15}.

		\begin{table}[]
		\centering
		\caption{Dielectric properties of different materials at frequency $39$ GHz.}
		\label{tb:DiElec}
			\resizebox{\columnwidth}{!}{
		\begin{tabular}{|l|l|l|l|}
			\hline
			\rowcolor[HTML]{EFEFEE} 
			Material &   Conductivity [S/m]     &  Relative permittivity  & Loss tangent       \\ \hline
			Skin (Dry) & 31.429 &	11.983&	1.2089  \\ \hline
			Skin (Wet) & 32.432 &	14.386&	1.0391 \\ \hline
			Muscle & 42.501 &	18.639 &	1.051  \\ \hline
			Fat & 2.174 &	3.424&	0.2926 \\ \hline
			Bone & 6.28 &	4.7268&	0.5637  \\ \hline
			Layer 1 & 31.9305 &	13.1845 &	1.124 \\ \hline
			Layer 2 & 12.703 & 7.248 &0.6069  \\ \hline
		\end{tabular}}
	\end{table}

    Given this model for the hand, we can show the effect of the hand blockage on the antenna's radiation pattern. In Fig. \ref{fig:RadWOandW}, we plot the spherical gain of one patch antenna element designed by us with and without blockage. Note that these results are directly taken from HFSS simulations. The figure shows about 20-25 dB loss in the boresight direction (z-axis). It also shows the irregular effect of the blockage on the radiation pattern which makes it hard to predict using an analytical model and justifies our reliance on HFSS simulations to capture the effect of reflections, couplings, and attenuation caused by fingers. 
    
	\subsection{Simulating Different Hand Grips}\label{Sec:SDHG}
	
	Due to the diverse hand grips in terms of the position of the fingers on the phone, simulating all of them, where we take precisely the positions of the fingers on the phone, is infeasible. Hence, we consider a simplified way where we partition the faces of the phone into disjoint regions, each with the size of the antenna module. We focus on the cases where the region is fully blocked. In other words, in case of blocking, we assume that the finger covers the whole region. This is a reasonable assumption given that the antenna modules are small compared to the finger due to the short wavelength.
	
	Moreover, since the loss is more severe when the fingers are in close proximity to the antenna, due to the reflections, coupling, and attenuation as we mentioned earlier, we focus on the regions at the corners of the mobile phone, where the antenna modules are placed. Hence, in total, we have $9$ possible regions, each corresponds to one of the modules shown in Fig. \ref{fig:Designs}. Note that since we consider only the the $9$ regions directly located above the modules, we neglect the cases where the human finger is placed on the face of the phone. However, due to the high attenuation of the screen, as shown in Fig. \ref{fig:RadWOandW}, the attenuation caused by the finger on these regions can be neglected.
	
	Based on this, we have $2^{9}$  possible combinations of blocked regions. Simulating all of these is infeasible since simulators like HFSS are time-consuming. Consequently, we simulate the $9$ cases which correspond to the presence of the finger on each module separately; these cases are referred to as the elementary cases hereafter. For each elementary case, we have an antenna response vector for each point $x_i$, $\mathbf M_i^{(j)}$, where $j\in \{1, 2, \cdots, 9\}$, which we get from HFSS. Then for the cases where more than one module is blocked, we construct the radiation pattern using the elementary cases as follows.
	
	Assume that for a certain grip $n$, the blocked modules are given by the set $\mathcal B \subset \{1, 2, \cdots, 9\}$ and the antenna response vector for this grip at point $x_i$ is denoted as $\mathbf M^{(n)}_i$. Then the response for the $k^{\rm th}$ antenna at point $x_i$ is given by
	\begin{equation} \label{Eq:elementPatternHandgrip1}
	\left[\mathbf M^{(n)}_i \right]_k= \left[\mathbf M^{(j^{*})}_i \right]_k,
	\end{equation}
	where,
	\begin{equation}\label{Eq:elementPatternHandgrip2}
	j^{*}= \argmin_{j \in \mathcal B} \left|\left[\mathbf M^{(j)}_i \right]_k \right|,
	\end{equation}
	where $\left[\mathbf{A}\right]_k$ stands of the $k^{\rm th}$ element of a vector $\mathbf{A}$.
	Eq. \eqref{Eq:elementPatternHandgrip1}-\eqref{Eq:elementPatternHandgrip2} mean that for each antenna and each point $x_i$, the antenna response is chosen to be the one that has the minimum gain from all the responses of the cases given in $\mathcal B$. 
	By following this approach, we can find the radiation pattern given any hand grip using just the $9$  elementary cases we discussed, which significantly simplifies finding the radiation patterns, since it can be done without the need of using HFSS.
	

\section{Activities and Hand Grips} \label{Sec:HandGrips}
	
\begin{table}[]
	\centering
	\caption{Different hand grips along with the blocked regions.}
	\label{tb:DiffGrips}
		\begin{tabular}{|l|l|l|l|}
			\hline
			\rowcolor[HTML]{EFEFEE} 
			Grip ID &   Blocked Modules  &Grip ID &   Blocked Modules     \\ \hline
			1 &  None & 8 & 1, 2, 3, 5, 7, 9.  \\ \hline
			2 & 1, 3, 4, 5, 7,9. &9 &   4, 7, 8, 9. \\ \hline
			3 &   7, 8, 9. &10 & 1, 2, 3, 4, 5, 7.\\ \hline
			4 &    7, 8.& 11 & 1, 2, 3, 5, 6, 7, 8, 9.\\ \hline
			5 &   8.&12 &  2, 3, 4, 5, 6.  \\ \hline
			6 & 1, 2, 3, 4, 5, 7, 9.&13 & 1, 2, 3.\\ \hline
			 7 & 1, 2, 3, 4, 5, 7, 8, 9. &14 & 1, 4, 7.\\ \hline
	\end{tabular}
\end{table}

	 To the best of our knowledge, there is no available database that records how users hold their phones while performing certain activities. Hence, we performed our own experiment. We asked eight users to hold a phone and  perform certain activities and then the blocked modules, shown in Fig. \ref{fig:Designs}, were noted by covering the phone with stickers.  The users were asked to watch a video, play a game, make a phone call, and text a message in both landscape and portrait orientations. The obtained hand grips are summarized in Table \ref{tb:DiffGrips}, where we list the blocked modules for each grip. For comparison, we include the no-blockage case and denote it as Grip $1$. In Table \ref{tb:ActGrips}, we show the type of grips observed for each activity along with their approximate frequencies. In addition, we have included the case where the phone is in the user's pocket and the screen is facing outwards for comparisons.

\begin{table}[]
	\centering
	\caption{Activities and hand grips.}
	\label{tb:ActGrips}
	{%
		\begin{tabular}{|l|l|l|}
			\hline
			\rowcolor[HTML]{EFEFEE} 
			Activity & Grip IDs & Respective Probabilities \\ \hline
			Game Portrait &  3 & $1$\\ \hline
			Game Landscape &  6, 7 ,8 & $\nicefrac{3}{8}, \ \nicefrac{2}{8},\  \nicefrac{3}{8}.$\\ \hline
			Video Portrait &  4, 9, 12 & $\nicefrac{3}{4}, \ \nicefrac{1}{8}, \ \nicefrac{1}{8}.$\\ \hline
			Video Landscape &  3, 10, 11, 12 & $\nicefrac{1}{8}, \  \nicefrac{3}{8}, \ \nicefrac{1}{4}, \ \nicefrac{1}{4}.$\\ \hline
			Messaging Portrait &  3 & $1$\\ \hline
			Messaging Landscape &  2, 6, 13 & $\nicefrac{3}{8}, \ \nicefrac{1}{4}, \ \nicefrac{3}{8}$\\ \hline
			Voice Call & 3, 4, 5& $\nicefrac{1}{4}, \ \nicefrac{1}{2}, \ \nicefrac{1}{4} $\\ \hline
			Pocket  & 14& 1\\ \hline
	\end{tabular}}
\end{table}

\section{Codebook Adaptation with Hand Grips}\label{Sec:Codebook}

Using HFSS along with our models for the mobile phone and the hand, we can obtain the antenna gains for each grip in Table \ref{tb:DiffGrips}. This section focuses on how the codebook design can be adapted with these grips.

\subsection{Grip-Aware Scheme}

 For this scheme, we assume that the phone knows exactly the hand grip and the corresponding antenna radiation patterns. With this knowledge, the phone determines a codebook that is specifically optimized for the current user grip. Hence, the codebook design algorithm in this scheme is the same as designing the codebook for the free space case described in Section \ref{Sec:SCaH}, except that the antenna response vectors $\mathbf M_i, \ \forall i \in \{1, \cdots N_p \}$ are replaced by the antenna response vectors for each grip which are found by using the methods we described in Section \ref{Sec:SDHG}. In total, we have a codebook specifically designed for each hand grip.
 
{ Although estimating the user hand grip by the mobile phone, through capacitive touch sensors and infrared proximity sensors \cite{Sensing_Hinckley00,Hand_Kim06}, can bring many benefits in terms of providing a natural experience between the mobile phone and the user, current devices do not have the capabilities to provide an accurate estimation of the user hand grip. Moreover, it requires a different codebook for each different grip which may not be feasible due to storage limitation in the RF chipset and the switching overhead. Accordingly, this scheme can be thought of as an upper bound on the gain we can achieve by adapting the codebook based on the user grip.}

\subsection{Semi-Aware Scheme}

 In this scheme, only the orientation of the phone and the application the user is using are assumed to be known by the mobile device. The rationale behind it is that although users hold their phones in various ways while performing the same activity depending on their personal habits and the environment, there are certain patterns that are highly correlated for a given activity. An analogy was drawn in \cite{Hand_Kim06} with traditional hand tools (e.g., hammer, cup, etc.), where the grip can be different for individual users, but still there is a correlation between different grips.\footnote{Note that in our work, we try to estimate the grip based on the activity and the orientation. In \cite{Hand_Kim06}, it is the opposite, i.e., they tried to estimate the activity based on the grip.}  Hence, in this scheme, we exploit this correlation by designing a codebook for each activity, instead of a codebook for each hand grip, which significantly reduces the number of stored codebooks. Moreover, current mobile phones can distinguish the orientation of the phone along with the application the user is using (or at least the genre of the activity).

The codebooks are designed as follows: for each pair of activity and orientation, design a codebook that maximizes the weighted mean of the spherical coverage over the common hand grips for this activity and orientation.  We choose the {\it weighted} mean since some grips are less common than others. However, the weighted mean is just an example of the desired objective function. The same algorithm can be used to maximize the minimum over all the grips (i.e., the design is based on the worst case scenario) or the sum-log of the spherical coverage, to ensure some fairness across different grips. 

Mathematically, denote the set of grips for an activity as $\mathcal B$ and the corresponding likelihood vector as ${\mathbf P}$. Then the codebook is found by solving the following optimization problem, which is an extension to \eqref{Eq:OptiPro}.
\begin{equation}
{\mathcal W_{c}}=\argmax_{ \left\{ {\mathbf w_1}, \cdots, {\mathbf w}_{N_c} \right\}\subset {\mathcal W_d}} \sum\limits_{j\in \mathcal B}{\mathbf P}(j) \bar{\mathcal{S}}^{(j)}(\{{{\mathbf w}_1, \cdots, {\mathbf w}_{N_c}}\}),\label{Eq:OptiProGrip}
\end{equation}
where $\bar{\mathcal{S}}^{(j)} (\cdot)$ is the average spherical gain, as defined in \eqref{Eq:AveSC}, for the $j^{\rm th}$ grip.

\subsection{Grip-Agnostic Scheme}
 This is the benchmark scheme, where the codebook is designed assuming no blockage, and the mobile phone does not adapt its codebook with different hand grips. Hence, the codebook is designed exactly as described in Section \ref{Sec:SCaH}.
 
 In the next section, we compare these schemes, and show the importance of including the hand grips into the beam codebook design.

\section{Simulation Results}\label{Sec:SimRes}
In this section, we study the performance of the different codebook adaptation schemes using the data presented in Section \ref{Sec:HandGrips}.  The raw radiation data is taken from HFSS and processed in MATLAB. For the spherical coverage, we focus on the region described by $0^\circ \leq \theta \leq 100^\circ$ and  $0^\circ\leq \phi <360^\circ$, where $\theta$ is  the elevation  angle, and $\phi$ is the azimuth angle, which includes the whole hemisphere facing the back of the phone, plus the adjacent $10$ degrees from the other hemisphere. The reason behind this choice is that the region directly facing the screen is already dead because of the screen blockage and the front-to-back attenuation of the patch antennas as we described earlier. Hence, changing the codebook design algorithms or the hand grip adaptation schemes will not have an impact on this region. The following values of the parameters are chosen: $N_p=5809$, $N_c=15$, and $N_d=363$.

\begin{table*}[t]
  \centering
  \caption{The gain in  the percentiles of the spherical coverage for different codebook adaptation schemes relative to the grip-agnostic scheme.}
    \begin{tabular}{|c|r|r|r|r|r|r|}
    \hline\rowcolor[HTML]{EFEFEE}
          & \multicolumn{2}{c|}{20$^{\rm th}$ Percentile} & \multicolumn{2}{c|}{50$^{\rm th}$ Percentile} & \multicolumn{2}{c|}{80$^{\rm th}$ Percentile} \\
    \hline\rowcolor[HTML]{EFEFEE}
    Activity & \multicolumn{1}{c|}{Semi-Aware} & \multicolumn{1}{c|}{Grip-Aware} & \multicolumn{1}{c|}{Semi-Aware} & \multicolumn{1}{c|}{Grip-Aware} & \multicolumn{1}{c|}{Semi-Aware} & \multicolumn{1}{c|}{Grip-Aware} \\
    \hline
    Call  & 26\%   & 27\%   & 19\%   & 19\%   & 8\%   & 10\% \\
    \hline
    Game Port & 37\%   & 37\%   & 21\%   & 21\%   & 12\%   & 12\% \\
    \hline
    Game Land & 34\%   & 49\%   & 44\%   & 54\%   & 25\%   & 30\% \\
    \hline
    Video Port & 20\%   & 29\%   & 16\%   & 18\%   & 11\%   & 11\% \\
    \hline
    Video Land & 20\%   & 29\%   & 16\%   & 18\%   & 11\%   & 11\% \\
    \hline
    Msg Port & 37\%   & 37\%   & 21\%   & 21\%   & 12\%   & 12\% \\
    \hline
    Msg Land & 13\%   & 23\%   & 7\%   & 15\%   & 10\%   & 18\% \\
    \hline
    Pocket & 57\%   & 57\%   & 40\%   & 40\%   & 26\%   & 26\% \\
    \hline
    \end{tabular}%
  \label{tab:MainRes}%
\end{table*}%

The 20$^{\rm th}$, 50$^{\rm th}$, and 80$^{\rm th}$ percentiles of the spherical coverage for the different codebook adaptation schemes are presented in Table \ref{tab:MainRes}. The results are found as follows.

For the {\it Grip-Agnostic} scheme, the codebook is designed based on the radiation vectors of the no blockage case. 
Then the performance of this codebook is evaluated for each hand grip in Table \ref{tb:DiffGrips}, in terms of the spherical coverage. Hence, we have 14 different sets of data which represents the spherical coverage for each grip assuming the no blockage codebook. Then, for each activity, the weighted mean of the percentiles, based on Table \ref{tb:ActGrips}, is found. For the  {\it Grip-Aware} scheme, a codebook is designed for each hand grip, and then the spherical coverage is found for each grip assuming the codebook that is specifically designed for it. The shown percentiles for the different activities are also the weighted mean of the percentiles for each grip. Finally, for the {\it Semi-Aware} scheme, a single codebook is designed for each activity as described in Section \ref{Sec:Codebook}. Then for each activity, the spherical coverage distributions are evaluated for the activity's corresponding grips assuming the codebook designed for this activity. The percentiles are found in a similar way to the previous cases. To simplify the comparison, the results are normalized by the {\it Grip-Agnostic} data, i.e., the results are the gains compared to the {\it Grip-Agnostic} scheme.

First, we start by comparing the {\it Grip-Aware} scheme, which represents the maximum gain we can obtain by adapting the codebook according to the hand grip, and the {\it Grip-Agnostic} scheme, which is our benchmark scheme that is based on the no-blockage case. As the figures show, we get a considerable gain that ranges between $23\%$ to $57\%$ at $20^{\rm th}$ percentile by adapting the codebook according to the user hand grip. We also see gains at the $50^{\rm th}$ and the $80^{\rm th}$ percentiles. However, an improvement of the $20^{\rm th}$ percentile is more important for coverage, since it corresponds to the regions where we have a low signal quality, and the user might have a link-failure in case the signal comes from these regions. Additionally, note that the gap between the {\it Grip-Aware} and the {\it Grip-Agnostic} schemes varies depending on the activity. In general, the more elements are blocked by the grips, the more benefits brought by adapting the codebook. Hence, for the activities that have grips blocking many elements, the gap between the two schemes is larger. These results also show that the designer does need to take the hand grip into account while designing the beam codebook. A careful codebook design by taking the hand grips into account is necessary and results in a significant performance gain.

However, the obtained gains from the {\it Grip-Aware} scheme may be unrealistic given the current mobile devices, since it is based on the assumption that the mobile device can accurately detect the user hand grip and this feature is not fully available in the current mobile phones. Moreover, the beam codebook switching based on the grip may result in high overheads that reduce the gain obtained from this scheme. Hence, we compare the {\it Grip-Aware} scheme with the proposed {\it Semi-Aware} scheme, since it is a practical scheme that only requires the knowledge of the application the user is using and the orientation of the phone. Moreover, since the same codebook is used for each activity, the switching overheads mentioned previously are significantly reduced. 

The results show that the {\it Semi-Aware} scheme provides gains ranging between $13\%$ and $57\%$ in terms of the $20^{\rm th}$ percentile, which are less compared to the gains offered by the {\it Grip-Aware}, but are still significant compared to the  {\it Grip-Agnostic} scheme. The variation in the gap between the {\it Semi-Aware} and the {\it Grip-Agnostic} schemes can be explained in the same way we explained the gap between the {\it Grip-Aware} and the {\it Grip-Agnostic} schemes. The variation in the gap between the {\it Grip-Aware} and the  {\it Semi-Aware} schemes can be explained by the grip correlation of a activity, i.e., the gap is smaller for the activities that have highly correlated grips. Of course, in the cases where there is only one grip, as in the case of the {\it Msg Portrait}, {\it Game Portrait}, and {\it Pocket} activities, the performance of the two schemes is exactly the same. Conversely, for the case of the {\it Call} activity, the performance is very close due to the high correlation between the different grips obtained for the Call activity as shown in Table \ref{tb:ActGrips}.  Overall, the results show that we can still harvest performance gains with just knowledge of the application and the mobile orientation.

\section{Conclusion} \label{Sec:Conc}
In this work, we studied how the mmWave beamforming codebook can be adapted based on the user handgrip. We consider a practical design of the antenna placement on the mobile phone and a practical codebook design scheme. Then we study three different codebook adaption schemes: a grip-aware scheme, where perfect knowledge of the hand grip is available; a semi-aware scheme, where just the application (voice call, messaging, etc.) and the orientation of the mobile handset is known; and a grip-agnostic scheme, where the codebook ignores hand blockage.  Our results show that the ideal grip-aware scheme and the practical semi-aware can provide up to $50\%$ gain in terms of the spherical coverage over the agnostic scheme, depending on the grips and user activities.

 \bibliographystyle{IEEEtran}
\bibliography{AhmadRefCodebook}
\vfill
\end{document}